\documentclass[prd,aps]{revtex4}
\usepackage{graphicx}
\def\ut#1{\rlap{\lower1ex\hbox{$\sim$}}#1{}}

\begin{document}
\title{Unified model of loop quantum gravity and matter}
\author{Rodolfo Gambini$^{1}$, S. Jay Olson $^{2}$ and Jorge Pullin$^{2}$}
\affiliation {1. Instituto de F\'{\i}sica, Facultad de Ciencias,
Igu\'a 4225, esq. Mataojo, Montevideo, Uruguay. \\
2. Department
of Physics and Astronomy, Louisiana State University,
Baton Rouge, LA 70803-4001}
\date{September 7th 2004}

\begin{abstract}
We reconsider the unified model of gravitation and Yang--Mills
interactions proposed by Chakraborty and Peld\'an, in the light
of recent formal developments in loop quantum gravity. In particular,
we show that one can promote the Hamiltonian constraint of the
unified model to a well defined anomaly-free quantum operator
using the techniques introduced by Thiemann, at least for the 
Euclidean theory. The Lorentzian version of the model can
be consistently constructed, but at the moment appears to yield
a correct weak field theory only under restrictive assumptions,
and its quantization appears problematic.
\end{abstract}

\maketitle

\section{Introduction}

The ``quantum geometry'' approach to quantum gravity
(also known as ``loop quantum gravity'') is based on the philosophy
that one wishes to understand in detail the difficulties that arise
when quantizing general relativity. Within this context, the issue
of unification with other interactions has not been considered too
frequently. Chakraborty and Peld\'an \cite{ChaPe} proposed some years
ago one possibility for unification. Their proposal is based on the
following observation. Consider first the constraints of general relativity
written in terms of Ashtekar's new variables,
\begin{eqnarray}
D_a \tilde{E}^{ai}&=&0, \label{gauss}\\
\tilde{E}^a_i F_{ab}^i&=&0,\label{diffeo}\\
\tilde{E}^a_i \tilde{E}^a_jF_{ab}^k \epsilon^{ijk}&=&0.\label{jamon}
\end{eqnarray}
At the moment we are referring to the original formulation of
Ashtekar's new variables, where the densitized triads $\tilde{E}^{ai}$
and the connection $A_a^i$ that defines the covariant derivative $D_a$
and the curvature $F_{ab}^i$ are in principle complex $SU(2)$ valued
quantities, and one needs to impose reality conditions to recover
ordinary real general relativity. Chakraborty and Peld\'an propose
generalizing these equation by considering that the variables take 
values in a more general gauge group. For simplicity, let us consider
such group to be $SU(N)$. Now, equations (\ref{gauss},\ref{diffeo}) 
are valid in general for any gauge group. The Hamiltonian constraint
(\ref{jamon}) requires some further elaboration. In particular one
needs a suitable generalization of the invariant tensor $\epsilon_{ijk}$.
The proposal consists in considering the following object,
\begin{equation}
\epsilon_{ijk}(\tilde{E}) = {\tilde{E}^a_i \tilde{E}^b_j \tilde{E}^c_k \ut{\eta}_{abc} \over
{\rm det} q}\label{epsilon}
\end{equation}
where by ${\rm det} q $ we mean the determinant of the spatial metric
$q_{ab}$,
\begin{equation}
({\rm det}q)q^{ab} =  \tilde{E}^a_i\tilde{E}^b_i.
\end{equation}

It is straightforward to show that when the gauge group is $SU(2)$ the
above expression reduces to the usual Levi--Civita invariant tensor,
and the spatial metric corresponds to the usual definition. If one
considers a more general gauge group, like $SU(N)$, the
$\epsilon(\tilde{E}) $ transforms covariantly under the group action,
but in general will be a function of the triads. Also the ``spatial
metric'' is clearly a generalization.  The resulting theory is a
different theory than general relativity.

Chakraborty and Peld\'an were able to show that if one uses the
proposed invariant tensor in the definition of the Hamiltonian
constraint there exist suitable limits in which the theory can, for
given gauge groups, approximate general relativity coupled with a
Yang--Mills field. They have also shown that the resulting constraint
algebra is first class.

At the time this proposal was presented, there were important
difficulties unresolved. In particular it was not clear how to
implement in the theory reality conditions that would guarantee that
the resulting theory was real general relativity coupled to fields. An
alternative was to consider the theory in a Euclidean signature and
then all variables were real. But there was no straightforward answer
in the case of a Lorentzian signature. Moreover the Hamiltonian
(\ref{jamon}) is non-polynomial in the basic variables, which seemed
a significant obstacle for quantization at the time.

New developments have taken place since this proposal. In particular,
Thiemann \cite{Th} has shown how to deal with non-polynomial terms in
the Hamiltonian constraint, including the ones that arise when one
considers the theory formulated in terms of real variables. In this
note we would like to show that these developments can be applied to
promote the Hamiltonian constraint of the Chakraborty and Peld\'an
proposal to a well defined quantum operator, at least for the Euclidean
theory.

\section{Making the (Euclidean) Hamiltonian constraint polynomial}

In the context of usual quantum geometry, Thiemann noticed the 
following important classical identity between the basic fields of 
the theory,
\begin{equation}
{\tilde{E}^a_i \tilde{E}^b_j \over \sqrt{\rm det}q }= 2 \left\{A_c,V\right\}^k \epsilon_{ijk} \eta^{abc},
\end{equation}
where $V$ is the volume,
\begin{equation}
V = \int d^3x \sqrt{\tilde{E}^a_i\tilde{E}^b_j\tilde{E}^c_k \epsilon^{ijk} \ut{\eta}_{abc}}.
\end{equation}
The importance of the identity introduced by Thiemann is that it
allows to express a non-polynomial expression in terms of an
expression involving a Poisson bracket and polynomial dependences,
which can be quantized straightforwardly.

We now proceed to generalize the above expression using the ideas of
Chakraborty and Peld\'an. We replace the $\epsilon$ by equation
(\ref{epsilon}) and let the internal gauge group be different from
$SU(2)$. The volume becomes,
\begin{equation}
V = \int d^3x \sqrt[4]{\tilde{E}^a_i\tilde{E}^d_i
\tilde{E}^b_j\tilde{E}^e_j\tilde{E}^c_k
\tilde{E}^f_k  \ut{\eta}_{abc}\ut{\eta}_{def}}.
\end{equation}

From here we can derive the identity,
\begin{equation}
2\left\{A_a,V\right\}^i = 
{ \tilde{E}^d_i\tilde{E}^b_j\tilde{E}^e_j\tilde{E}^c_k
\tilde{E}^f_k  \ut{\eta}_{abc}\ut{\eta}_{def} \over ({\rm det} q)^{3/2}}.
\end{equation}

Remarkably, we can use this latter identity to write the (Euclidean)
Hamiltonian constraint in the generalized case as,
\begin{equation}
H(N) =\int d^3 x  \left\{A_c,V\right\}^k F_{ab}^k \tilde{\eta}^{abc},
\end{equation}
and surprisingly, the Hamiltonian takes the same form as the one
Thiemann proposed for general relativity. From here on one could
proceed to the quantization using the same tools Thiemann used in the
usual gravitational case. The generalization is straightforward, one
would consider  wavefunctions that are given by spin networks of the
appropriate gauge group. It is remarkable that the generalization
proposed by Chakraborty and Peld\'an acquires such a simple form in
terms of spin networks.

One element we need to verify is the fact that the volume operator we
are using does not take the standard form. Can we be sure it is a well
defined operator? The easiest way to see this is to look at
the version of the proof of finiteness of the volume operator
presented by Ashtekar and Lewandowski \cite{AsLe}. They use strip
operators to represent the triads, that correspond to triads smeared
along a two dimensional surface. When one acts with the triads on a
spin network state one generates a contribution proportional to the
Dirac delta integrated along the spin net. That integral, together
with the two dimensional integral of the smearing, yields a finite
result when integrating the three dimensional Dirac delta. The same
argument would apply to the operator we are presenting here.

\section{The real Lorentzian theory}

Up to now we considered the Hamiltonian constraint that would arise
with real variables in the Euclidean theory. Let us analyze the
Lorentzian case.  Ashtekar introduced the new variables for canonical
gravity by starting with the usual formulation in terms of
(densitized) triads $\tilde{E}^a_i$ and the canonical momentum
$K_a^i\equiv K_{ab}E^{ai}$ related to the extrinsic curvature
$K_{ab}$.  He then considers a canonical transformation given by
$A_a^i =i K_a^i+\Gamma_a^i$ where $\Gamma_a^i$ is the spin connection
compatible with the triad.  Barbero \cite{Ba} made explicit what would
happen if one considered a more general canonical transformation
$A_a^i =\beta K_a^i +\Gamma_a^i$ with $\beta$ a parameter usually
referred to as the Immirzi parameter. The Gauss law and diffeomorphism
constraint remain unchanged. The Hamiltonian constraint however,
becomes,
\begin{eqnarray}
H(N) &=& H^E(N)+H^{\rm Lor}(N)\\ 
H^E(N) &=&\int d^3x N(x) {\tilde{E}^a_i \tilde{E}^b_j \over \sqrt{\rm det q}} F_{abk}\epsilon^{ijk}\\
H^{\rm Lor}(N) &=& -\int d^3 x N(x)   2 {(\beta^2+1) \over \beta^2} 
{\tilde{E}^{[a}_i \tilde{E}^{b]}_j \over \sqrt{\rm det q}}K^a_i K^b_j,
\end{eqnarray}
where we have distinguished the piece $H^E(N)$ that by itself would be
the Hamiltonian constraint of the Euclidean theory if the variables
are real, or the Hamiltonian of the Lorentzian theory if $\beta=i$ and
the extra piece $H^{\rm Lor}$ that is needed to be included in order
to have the Lorentzian theory with $\beta$ a real quantity.

The choice of $\beta=i$ makes the second term vanish, and this was the
original motivation for choosing that value in the Ashtekar
formulation. If one chooses $\beta$ a real number, one needs to deal with the
second term. Thiemann also taught us how to do this. He starts by
noting that the trace of the extrinsic curvature satisfies,
\begin{equation}
K = -\left\{{V },\int d^3x H^E(x)\right\}
\end{equation}
where $H^E(x)$ is the single-densitized Euclidean Hamiltonian
constraint $H^E(N)=\int d^3x N(x) H^E(x)$ and then the extrinsic
curvature with one triad index can be written as,
\begin{equation}
K_a^i = \left\{{A_a^i},K\right\}.
\end{equation}

Finally, the extra portion of the Hamiltonian that arises in the
Lorentzian theory, can therefore be written as,
\begin{equation}
H^{\rm Lor}(N) = -\int d^3 x N(x)   2 {(\beta^2+1) \over \beta^2}
\left\{A_a^i,V\right\}
\left\{A_b^j,K\right\}
\left\{A_c^k,K\right\}\tilde{\eta}^{abc} \epsilon_{ijk}.
\end{equation}

The question now arises of how to generalize all this to
to $SU(N)$ using the
Chakraborty and Peld\'an idea. The particular form of $H^{\rm Lor}$ we
just mentioned may be problematic to generalize to $SU(N)$, since
it involves explicitly $\epsilon_{ijk}$. If one replaces this 
quantity using the dynamical epsilon of Chakraborty and Peld\'an,
it is not clear that the resulting constraint will close the
appropriate constraint algebra. If we ignore this point for just
a second, it is worthwhile noting that the resulting object is likely
to be quantizable  (see section 3.4 of
\cite{qsd5}), and given the general form of the expressions the techniques
of \cite{qsd5} produce, the resulting quantum operator may satisfy the 
correct commutator algebra, at least operating on diffeomorphism 
invariant functions.  An alternative to this is to write $H^{\rm Lor}$
without using the Poisson bracket tricks directly in terms of $\tilde{E}^a_i$
and $K_a^i$, the extrinsic curvature. The latter can be computed in terms
of the canonical variables. The resulting expression does not involve $\epsilon^{ijk}$ at all, so generalizing it to $SU(N)$ will produce an expression
that classically still satisfies the correct Poisson algebra.  However,
the techniques of \cite{qsd5} may not help in its quantization since they
involve the use of epsilons.

Unfortunately, in both choices one is further confronted with the
problem of reproducing Einstein's theory coupled to Yang--Mills using
the same assumptions as Chakraborty and Peld\'an, namely that one is
considering weak fields living on a DeSitter background. Without
further assumptions the extra portion of the Hamiltonian constraint
will yield terms coupling to the extrinsic curvature. One can
eliminate these terms and obtain Einstein--Yang--Mills theory by
choosing a slicing in which the extrinsic curvature vanishes, but this
appears rather unnatural.  This difficulty is rather fundamental,
since when one uses real Ashtekar variables the condition for the
background to be DeSitter space is more complicated than in the
Euclidean case, and it involves explicitly the extrinsic curvature.

Another outstanding issue with this attempt to unification is the
introduction of Fermions. This is due to the fact that in the action
for Fermionic fields the gauge group generators appear explicitly, and
key identities of $SU(2)$ play a crucial role in the calculations that
lead to showing that the constraint algebra closes. It is possible
that a different approach, for instance using supersymmetry, could be
used.  The Ashtekar formulation has been generalized to supergravity
\cite{sugra}, so an avenue of attack of this problem is available.
Scalar matter has been included in the model successfully by 
Chakraborty and Peld\'an.

\section{Summary}

Summarizing, we have shown that Thiemann's quantization technique can
be applied to the Chakraborty--Peld\'an unified model yielding a well
defined quantum theory at least for the Euclidean signature. It is
remarkable that the Euclidean portion of the Hamiltonian constraint is
exactly the same as in the usual case just acting on spin network
states of a more general gauge group.  The full Lorentzian theory
however has a Hamiltonian constraint that is more complicated than the
purely gravitational one and it is not clear how to naturally get a
limit in which it yields the Einstein--Yang--Mills theory.  If this
difficulty could be overcome and the model could be made to include
Fermions via supersymmetry it would become a viable, elegant and
mathematically well define way of having a unified theory of quantum
fundamental interactions.

\section{Acknowledgments}
We wish to thank Octavio Obreg\'on for discussions.  This work was
supported by grant NSF-PHY0244335 and funds from the Horace Hearne Jr.
Laboratory for Theoretical Physics.


\begin{thebibliography}{99}
\bibitem{ChaPe} S.~Chakraborty and P.~Peldan,
Phys.\ Rev.\ Lett.\  {\bf 73}, 1195 (1994) [arXiv:gr-qc/9401028];
Int.\ J.\ Mod.\ Phys.\ D {\bf 3}, 695 (1994) [arXiv:gr-qc/9403002].
\bibitem{Th} 
T.~Thiemann,
Class.\ Quant.\ Grav.\  {\bf 15}, 839 (1998)
[arXiv:gr-qc/9606089].
\bibitem{AsLe} A.~Ashtekar and J.~Lewandowski,
Adv.\ Theor.\ Math.\ Phys.\  {\bf 1}, 388 (1998)
[arXiv:gr-qc/9711031].
\bibitem{Ba} J.~F.~Barbero Phys.\ Rev.\ D {\bf 51}, 5507 (1995)
[arXiv:gr-qc/9410014].
\bibitem{qsd5}
T.~Thiemann,
Class.\ Quant.\ Grav.\  {\bf 15}, 1281 (1998)
[arXiv:gr-qc/9705019].
\bibitem{sugra}
T. Jacobson, Class. Quant. Grav.5(1988)923; 
D. Armand-Ugon, R. Gambini, O. Obregon, J. Pullin,   Nucl. Phys.
B460 (1996) 615 [arXiv:hep-th/9508036]; L. Urrutia, in ``Proceedings
of the 5th Mexican Workshop of particles and fields'' J. C. D'Olivo,
A. Fernandez, M. Perez, editors, AIP Press, Woodbury, NY 1996 
[hep-th/9609010];  H. Kunitomo and T. Sano
Prog. Theor. Phys. Suppl. {\bf 114}, 31 (1993); T. Sano and J. Shiraishi,
Nucl. Phys. {\bf B410}, 423 (1993) [arXiv:hep-th/9211104];  T. Kadoyoshi and S. Nojiri, Mod. Phys. Lett. {\bf A12}, 1165 (1997) [arXiv:hep-th/9703149];
K. Ezawa, Prog. Theor. Phys. {\bf 95}, 863 (1996) [arXiv:hep-th/9511047];
Y. Ling, J. Math. Phys. {\bf 43}, 154 (2002) [arXiv:hep-th/0009020];
Y.~Ling and L.~Smolin,Nucl.\ Phys.\ B {\bf 601}, 191 (2001)
[arXiv:hep-th/0003285]
\end{thebibliography}
\end{document}